\documentclass[conference,letterpaper]{IEEEtran}
\IEEEoverridecommandlockouts

\usepackage[ruled,linesnumbered]{algorithm2e}
\usepackage{amsmath, amssymb, amsfonts}
\usepackage{cite}
\usepackage{graphicx}
\usepackage{multirow}
\usepackage{stfloats}
\usepackage[caption=false]{subfig}
\usepackage{xcolor}
\def\BibTeX{{\rm B\kern-.05em{\sc i\kern-.025em b}\kern-.08em T\kern-.1667em\lower.7ex\hbox{E}\kern-.125emX}}
\begin{document}
\title{Dynamic Mask Enhanced Intelligent Multi-UAV Deployment for Urban Vehicular Networks
\thanks{This work is supported by  the National Natural Science Foundation of China (Grant No. U23A20275 and 62201543), and by the Fundamental Research Funds for the Central Universities (Grant No. WK2100250071).}
\thanks{H. He is the corresponding author.}
}

\author{
\IEEEauthorblockN{Gaoxiang Cao\IEEEauthorrefmark{1}, Wenke Yuan\IEEEauthorrefmark{1}, Yunpeng Hou\IEEEauthorrefmark{2}, Huasen He\IEEEauthorrefmark{1}\IEEEauthorrefmark{2}, Quan Zheng\IEEEauthorrefmark{1}\IEEEauthorrefmark{2}, and Jian Yang\IEEEauthorrefmark{1}\IEEEauthorrefmark{2}}
\IEEEauthorblockA{\IEEEauthorrefmark{1}Department of Automation, University of Science and Technology of China, Hefei, China \\
\IEEEauthorrefmark{2}Institute of Artificial Intelligence, Hefei Comprehensive National Science Center, Hefei, China\\
\{cgx, ywk4432, hyp314\} @mail.ustc.edu.cn, \{hehuasen, qzheng, jianyang\} @ustc.edu.cn}}

\IEEEoverridecommandlockouts
\IEEEpubid{
  \begin{minipage}{\textwidth}
    \vspace{40pt} 
    \centering
    \scriptsize
    \copyright 2026 IEEE.  Personal use of this material is permitted.  Permission from IEEE must be obtained for all other uses, in any current or future media, including reprinting/republishing this material for advertising or promotional purposes, creating new collective works, for resale or redistribution to servers or lists, or reuse of any copyrighted component of this work in other works.
  \end{minipage}
}

\maketitle

\begin{abstract}
    Vehicular Ad Hoc Networks (VANETs) play a crucial role in realizing vehicle-road collaboration and intelligent transportation. However, urban VANETs often face challenges such as frequent link disconnections and subnet fragmentation, which hinder reliable connectivity.  To address these issues, we dynamically deploy multiple Unmanned Aerial Vehicles (UAVs) as communication relays to enhance VANET.  A novel Score based Dynamic Action Mask enhanced QMIX algorithm (Q-SDAM) is proposed for multi-UAV deployment, which maximizes vehicle connectivity while minimizing multi-UAV energy consumption. Specifically, we design a score-based dynamic action mask mechanism to guide UAV agents in exploring large action spaces, accelerate the learning process and enhance optimization performance. The practicality of Q-SDAM is validated using real-world datasets. We show that Q-SDAM improves connectivity by 18.2\% while reducing energy consumption by 66.6\% compared with existing algorithms.
\end{abstract}


\section{Introduction}

    With the development of smart cities and intelligent transportation systems, the Vehicular Ad Hoc Network (VANET) has become a key technology for vehicle-road collaboration and intelligent driving \cite{VANET_survey}. However, in complex urban environments, VANETs suffer from frequent link disruptions and dynamic topologies, hindering low-latency and reliable services. Untimely information transmission further affects traffic safety and efficiency, calling for complementary solutions. Recently, Unmanned Aerial Vehicles (UAVs), with their mobility, flexible deployment, and line-of-sight (LoS) communication advantages \cite{UAV_survey}, have been proposed as effective complements to VANETs, serving as aerial base stations or relays to mitigate disconnections and adapt to varying traffic.
    
    Deploying UAVs to enhance VANETs faces multiple challenges. The continuous movement for blind spot coverage drains energy, while reduced mobility lowers efficiency. In addition, multi-UAV collaboration requires intelligent scheduling to avoid redundancy. Moreover, UAVs must merge fragmented subnetworks to extend coverage. This multi-UAV deployment requires long-term planning, with computational scale increasing exponentially with the number of UAVs, making conventional algorithms struggle to solve. Recent studies show that Deep Reinforcement Learning (DRL) \cite{QMIX, IPPO} can effectively address such challenges.

\subsection{Related Work}

    Existing studies have made progress in UAV-aided VANETs. The authors in \cite{rel1} investigated data transmission in UAV-assisted VANETs and proposed a BPSO-based cooperative caching scheduling algorithm. By integrating micro UAVs into bandwidth-intensive and latency-sensitive applications, communication architectures in urban VANETs were explored in \cite{rel2}. The authors in \cite{rel3} designed a ADMR protocol to enhance communication reachability between UAVs and vehicles. A DACR algorithm was proposed in \cite{rel4}, which incorporated IoT elements into VANETs to improve connectivity. 

   Existing research has also made considerable progress in applying DRL for intelligent decision-making in UAV-assisted networks. The authors in \cite{rel5} employed the Soft Actor-Critic algorithm to maximize Quality of Experience (QoE) while reducing UAV flight energy consumption for real-time video transmission. The DRL based joint Trajectory Control and Offloading Allocation (DRL-TCOA) algorithm for UAV-assisted VANET edge computing was designed in \cite{rel6}, which significantly reduced system costs while ensuring the timeliness of task offloading. Adaptive Artificial Fish Swarm Algorithm driven Double Deep Q-Network (AAFSA-DDQNet) was proposed in \cite{rel7} to optimize dynamic resource allocation in VANETs. The authors in \cite{rel8} introduced a federated DRL-based cooperative offloading strategy that improved load balancing while ensuring task processing latency.

\subsection{Contribution}

    The limited coverage of Road Side Units (RSUs) and non-uniform distribution of vehicles make it challenging to maintain connectivity of VANET. To address this gap, we aim to enhance VANET connectivity through the dynamic deployment of UAVs as communication relays. Nevertheless, existing DRL-based methods struggle to handle large action space, which reduces the efficiency of policy exploration, limiting the optimization of VANET connectivity performance. Meanwhile, a large action space amplifies the variance of policy gradient estimation, leading to oscillation in learning. To cope with these challenges, we propose a score-based dynamic action masking algorithm to effectively guide agent exploration during training phase, accelerating the learning process.
    Our contributions are summarized as follows:

    \begin{itemize}
        \item We propose a VANET connectivity quantification method based on Road Topology Graph (RTG) and Dual RTG (DRTG), and formulate the dynamic deployment of multiple UAVs as an optimization problem for enhancing VANET connectivity in urban scenarios.
        \item We propose a Score based Dynamic Action Mask enhanced QMIX (Q-SDAM) algorithm for dynamically deploying multiple UAVs to enhance VANET connectivity. This algorithm incorporates a Score based Dynamic Action Mask (SDAM) mechanism to guide UAV agents in exploring the environment. As the number of UAVs increases, traditional algorithms struggle to cope with the exponential growth of the decision space, while SDAM's directional guidance effectively addresses this challenge.
        \item We evaluate the algorithm's performance using real-world datasets, demonstrating that Q-SDAM improves connectivity by 18.2\% while reducing energy consumption by 66.6\% compared with existing algorithms.
    \end{itemize}

    The remainder of this paper is organized as follows. Sec. \ref{sec:system_model} introduces the proposed system model and formulates the optimization problem. Sec. \ref{sec:proposed_approach} presents the proposed Q-SDAM algorithm in detail. Sec. \ref{sec:performance_evaluation} constructs a simulation system based on real-world datasets and discusses the simulation results. Sec. \ref{sec:conclusion} concludes the paper.

\section{System Model}\label{sec:system_model}
    
\begin{figure}[ht]
        \centering
        \includegraphics[width=.8\linewidth]{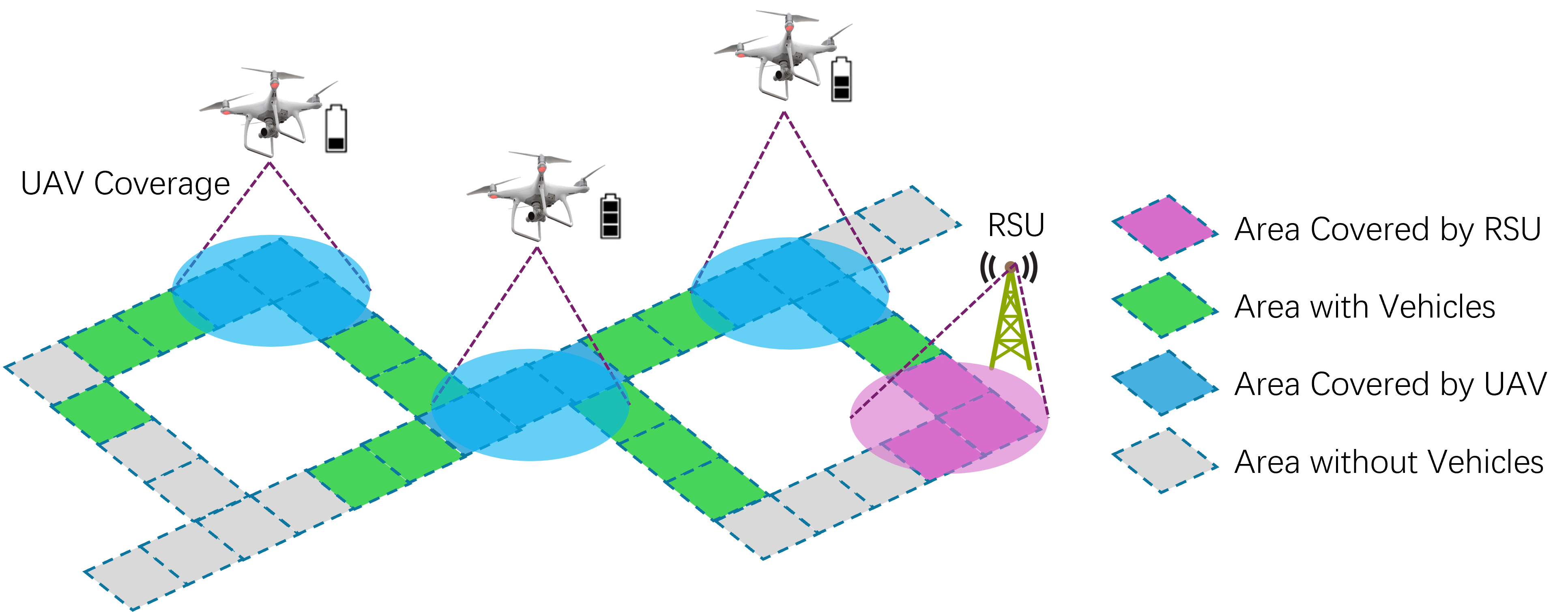}
        \caption{Multi-UAV Enhanced VANET Scenario}
        \label{fig:scenario}
    \end{figure}

    As illustrated in Fig. \ref{fig:scenario}, we consider a connected network of vehicles within a specific urban area over a certain period. Ground vehicles connect to each other via on-board wireless interfaces to form connected components, which are marked as green areas in the figure. Additionally, several RSUs are deployed in the city, consistently covering parts of the urban roads, denoted as purple areas in the figure. Due to the high mobility of vehicles, there remains disconnected regions between connected components (marked as gray areas in the figure), making it challenging to ensure the connectivity of VANET. To cope with this problem, multiple UAVs acting as communication relays are deployed to enhance the coverage of VANET, which are marked as blue areas in the figure. Air-to-Ground (A2G) links are employed for communication between UAVs and ground vehicles \cite{A2G}. The high line-of-sight probability on urban roads facilitates the establishment of communication connections between vehicles on the same road segment. Thus, we consider dynamically deploying $C$ UAVs at road intersections from the perspective of conserving UAV resources. This deployment strategy enables the UAVs to provide communication services for adjacent road segments. A discrete time-slot model consisting of $T$ time slots is adopted. It is assumed that UAVs take off from UAV service stations located at designated intersections and change their deployment positions at the start of each time slot. The set of UAVs is denoted as $\mathcal U = \{u_1, u_2, \cdots, u_C\}$.

\subsection{Graph Model for multi-UAV Assisted VANET}

    We assume that the road traffic network in the mission area comprises $n$ road intersections and $m$ ground roads connecting these intersections. Accordingly, an undirected graph with $n$ vertices and $m$ edges is constructed, where road intersections serve as vertices and ground roads act as edges. This graph is referred to as the Road Topology Graph (RTG), denoted as $\mathcal G = (\mathcal V, \mathcal E)$, where $|\mathcal V| = n$ and $|\mathcal E| = m$. The weight of a vertex $v$ is denoted as $p_v \in [-1, C]$, where $p_v = u$ $(u > 0)$ indicates that UAV $u$ is deployed at the road intersection corresponding to vertex $v$. $p_v = 0$ means the road intersection is consistently covered by an RSU. In contrast, $p_v = -1$ signifies that the road intersection is uncovered. The weight of an edge $e$ is denoted as $p_e$, representing the number of vehicles connected to the VANET on the road corresponding to edge $e$ during the current time slot. For the sake of convenience in description, no distinction is made between a vertex $v$ and its corresponding road intersection, nor between an edge $e$ and its corresponding urban road. It is assumed that the communication range of the UAV on-board wireless interface is sufficient to cover the length of any road segment. Consequently, any two vehicles on two different edges $e_1$ and $e_2$ connecting to the same vertex can communicate with each other. An edge with vehicles is termed as a \textit{c-edge}, and the vertices connected by c-edges are defined as \textit{c-vertices}. C-edges linked to the same c-vertex form a \textit{c-component}. Vertices covered by UAVs (temporarily) or RSUs (permanently) are specified as c-vertices, and the edges connected to these vertices are designated as c-edges.

\subsection{UAV Energy Consumption Model}

    The energy consumption of a generic UAV $u$ is modelled as flight energy consumption $e_u^f(t)$, hovering energy consumption $e_u^h(t)$, and communication energy consumption $e_u^c(t)$. Among these, $e_u^h(t)$ and $e_u^c(t)$ are proportional to the hovering time of the UAV, while $e_u^f(t)$ is proportional to the flight distance of the UAV \cite{DMTD}. Let $h_u(t)$ denote the hovering duration of UAV $u$ during time slot $t$, and let $l(v_i, v_j)$ represent the distance between vertex $v_i$ and vertex $v_j$ when the UAV flies from $v_i$ to $v_j$. Then, the total energy consumption of the UAV during this time slot is given by
    \begin{equation}
        \begin{aligned}
            e_u(t) =& e_u^f(t) + e_u^h(t) + e_u^c(t) \\
            =& (\varepsilon_1 + \varepsilon_2)h_u(t) + \varepsilon_3 l(v_i, v_j)
        \end{aligned}
    \end{equation}
    where $\varepsilon_1$, $\varepsilon_2$, and $\varepsilon_3$ are the energy consumption per unit.

\subsection{Dual Road Topology Graph}

\begin{figure}[ht]
        \centering
        \subfloat[Example RTG]{
            \includegraphics[width=.4\linewidth]{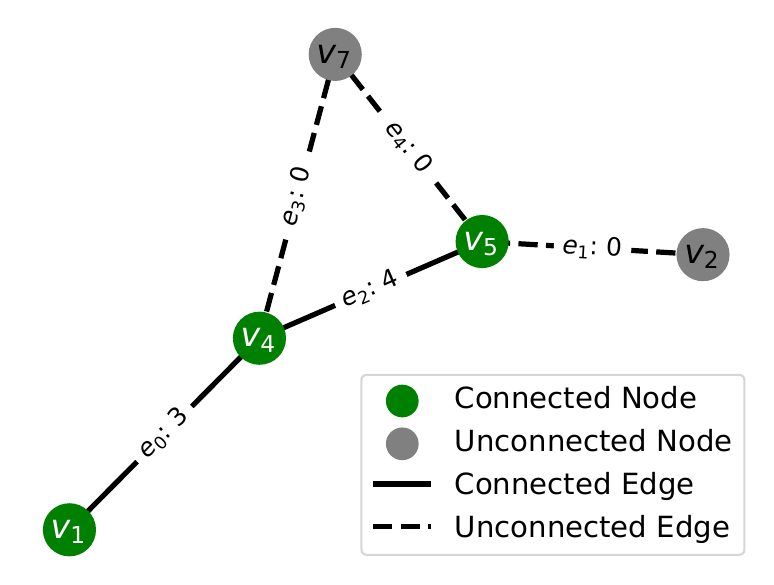}
            \label{fig:example_rtg}
        }
        \subfloat[Example DRTG]{
            \includegraphics[width=.4\linewidth]{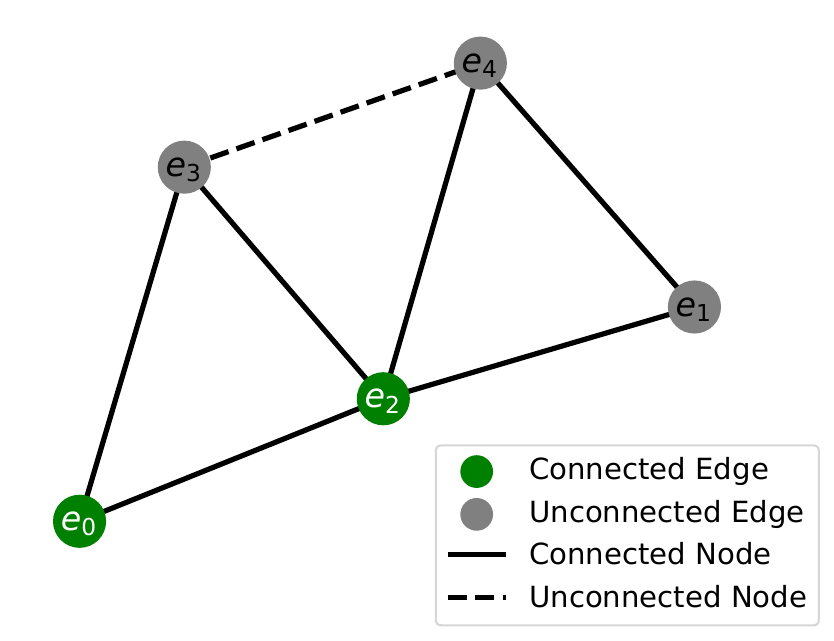}
            \label{fig:example_drtg}
        }
        \caption{Illustration of RTG and DRTG}
        \label{fig:example_graph}
    \end{figure}

    To facilitate description and formulate the optimization problem, we first introduce the Dual Road Topology Graph (DRTG) $\mathcal G^* = (\mathcal V^*, \mathcal E^*)$. For any edge $\forall e \in \mathcal E$, a unique vertex $v^* = \phi(e) \in \mathcal V^*$ is defined based on the bijective mapping $\phi: \mathcal E \to \mathcal V^*$. For any vertex $\forall v \in \mathcal V$ and any two edges $e_1, e_2 \in \delta(v)$ (where $\delta(v)$ denotes the set of edges incident to vertex $v$), the edge $(\phi(e_1), \phi(e_2)) \in \mathcal E^*$ is defined. The weight of a vertex $v^*$ in $\mathcal G^*$ is set to the weight of $\phi^{-1}(v^*)$ (the inverse mapping of $\phi$ at $v^*$), which corresponds to the number of vehicles connected to the VANET on the respective road. The vertices in $\mathcal V^*$ are colored according to the connection status of the edges in $\mathcal E$. Without loss of generality, c-vertices are colored black. In this case, the c-components mentioned earlier have a one-to-one correspondence with the connected subgraphs induced by the black vertex set in $\mathcal V^*$. The number of vehicles within a c-component is equivalent to the sum of the weights of the vertices in the corresponding connected subgraph. Taking Fig. \ref{fig:example_graph} as an example, Fig. \ref{fig:example_rtg} presents a simple RTG consisting of 5 nodes and 5 edges, while Fig. \ref{fig:example_drtg} shows the DRTG corresponding to this RTG. In the DRTG, ``$e_i: j$" indicates that there are $j$ vehicles connected to the VANET on edge $e_i$. C-vertices are represented as black vertices in Fig. \ref{fig:example_rtg} and correspond to solid edges in Fig. \ref{fig:example_drtg}. Other vertices are shown as gray vertices in Fig. \ref{fig:example_rtg} and correspond to dashed edges in Fig. \ref{fig:example_drtg}. The only c-component in Fig. \ref{fig:example_rtg} consists of edges $e_0$ and $e_2$ connected by the c-vertex $v_4$, with a total of 7 vehicles connected to the VANET. This corresponds to the only connected subgraph $\{e_0, e_2\}$ induced by the black vertex set in Fig. \ref{fig:example_drtg}, and the sum of the weights of the vertices in this subgraph is 7.

\subsection{Problem Formulation}
    \enlargethispage{-0.2in}
    To enhance the connectivity of VANET, we aim to maximize the number of vehicles within c-components. Let there be $k$ black connected subgraphs in the DRTG, with the sums of their vertex weights being $n_1(t), n_2(t), \cdots, n_k(t)$ respectively. We define the maximization optimization objective $\mathcal O_1$ as the average number of vehicles within the c-components, which is expressed as
    \begin{equation}
        \mathcal O_1 = \frac{1}{T \cdot k}\sum_{t=1}^T\sum_{i=1}^k n_i(t).
        \label{equ:O_1}
    \end{equation}
    On the other hand, due to the energy constraints of UAVs, we also aim to optimize the energy consumption of UAVs. The minimization optimization objective $\mathcal O_2$ is defined as the average energy consumption of UAVs per time slot, given by
    \begin{equation}
        \mathcal O_2 = \frac{1}{C \cdot T}\sum_{u=1}^C\sum_{t=1}^T e_u(t).
        \label{equ:O_2}
    \end{equation}
    
    In summary, the dynamic deployment problem of the multi-UAV-assisted VANET can be modelled as
    \begin{equation}
        \begin{aligned}
            & \max_{pos_u(t)}\ \mathcal O_1 - \mathcal O_2, \quad u \in [1, C], t \in [1, T] \\
            \text{s.t. } & 1.\ E_u(T) > 0, u \in [1, C] \\
            & 2.\ pos_u(t) \in \mathcal V, u \in [1, C], t \in [1, T]
        \end{aligned}
        \label{equ:optimization}
    \end{equation}
    where $pos_u(t)$ denotes the deployment position of UAV $u$ during time slot $t$, and $E_u(T)$ represents the remaining energy of UAV $u$ at the end of the mission.
    
    The optimization problem defined in \eqref{equ:optimization} is a complex Mixed-Integer Non-Linear Programming (MINLP) problem, which is difficult to solve using conventional optimisation algorithms \cite{discount}. Therefore, we leverage Multi-Agent Deep Reinforcement Learning (MADRL) techniques to efficiently solve this optimization problem. However, directly applying MADRL faces challenges such as large action spaces and inefficient exploration, which may hinder its performance. To address these issues, we design an enhanced framework, named SDAM, guiding the UAV agents' environment exploration, thereby enhancing their Performance.

\section{Proposed Algorithm}\label{sec:proposed_approach}

    In this section, we propose the DRL based Q-SDAM to optimize the cooperative deployment of multiple UAVs.  

\subsection{Key Elements of DRL}

    \textbf{Action:}  We define the action of the UAV $u$ in time slot $t$ as
    \begin{equation}
        a_u(t) = pos_u(t), \quad pos_u(t) \in \mathcal V.
        \label{equ:action}
    \end{equation}
    
    \textbf{Observation \& state:} We define the observation of the UAV $u$ in time slot $t$ as
    \begin{equation}
        o_u(t) = \{p_e(t),con_e(t) | e \in \mathcal E\} \cup \mathcal P_u(t),
    \end{equation}
    where $\mathcal P_u(t)$ is a one-hot vector with 1 at the $a_u(t)$-th position and 0 elsewhere, and $con_e(t)$ denotes the connection state of edge $e$ in time slot $t$. The system state at time slot $t$ is analogous to the agents' observation, except that the positions of the remaining UAVs in $\mathcal P_u(t)$ are also set to 1.
    
    \textbf{Reward function:} The reward of the multi-agent system in time slot $t$ is defined as
    \begin{equation}
        r(t) = \frac{\alpha_0}{\mu_1k}\sum_{i=1}^k n_i(t) - \frac{\beta_0}{\mu_2C} \sum_{u=1}^C e_u(t),
    \end{equation}
    where $\alpha_0$ and $\beta_0$ are reward weighting coefficients, and $\mu_1$ and $\mu_2$ are normalization coefficients.

    \begin{figure*}[ht]
        \begin{equation}
            f(a_u(t)) = \begin{cases}
                -1 - l(a_u(t), a_u(t - 1)) / \varphi_0, & \text{if}\ con(a_u(t)) = 1 \\
                - \dfrac{\alpha_1}{\varphi_0 } l(a_u(t), a_u(t - 1)) 
                + \dfrac{\beta_1}{\varphi_1} \sum\limits_{v \in N(a_u(t))} con(v), & \text{otherwise}
            \end{cases}
            \label{equ:score_func}
        \end{equation}
    \end{figure*}

\subsection{Score based Dynamic Action Mask}

    As shown in Eq. \eqref{equ:action}, the decision space size is $|\mathcal V|^C$, which grows exponentially with the number of UAVs, greatly increasing training cost and risking non-convergence. To address this, we propose a Score based Dynamic Action Mask (SDAM), which guides exploration and accelerates training.

    We define the action scoring function $f(\cdot): \mathcal V \to \mathbb R$ as Eq. \eqref{equ:score_func}, where $con(v, t)$ denotes the connectivity state of vertex $v$ at time slot $t$, $\alpha_1$ and $\beta_1$ are action scoring weights, and $\varphi_0$ and $\varphi_1$ are normalization parameters. When $con(a_u(t)) = 1$, $f(\cdot)$ prevents the UAV from moving to an already connected vertex, and if the UAV has to move to such a vertex, $f(\cdot)$ encourages it to go to the closest possible position to save energy. When $con(a_u(t)) = 0$, the first term encourages the UAV to prioritize vertices with shorter flight distances, while the second term encourages the UAV's target position to connect more c-components, thereby contributing to the optimization objective. When the agent selects an action, it calculates $f(\cdot)$ for $\{v_1, \cdots, v_n\}$ and sorts them to obtain a sequence $\{v_{k_1}, \cdots, v_{k_n}\}$ such that
    \begin{equation}
    f(v_{k_p}) \geqslant f(v_{k_q}), \forall 1 \leqslant p \leqslant q \leqslant n.
    \end{equation}
    The action mask is defined as a vector $\boldsymbol{\mu} \in \{0, 1\}^n$ whose components satisfy
    \begin{equation}
        \mu_i = \begin{cases}
            1, & i \in \{k_1, k_2, \cdots, k_{N_A}\} \\
            0, & \text{otherwise}
        \end{cases}.
    \label{equ:action_mask}
    \end{equation}
    The action mask $\boldsymbol{\mu}$ limits the number of available actions of the agent to $N_A$, guiding the agent's decision-making process.

    In the early stages of DRL algorithm training, $\boldsymbol{\mu}$ can effectively guide the agent to explore more efficient actions. However, as training progresses, due to the suboptimality designed based on the greedy algorithm, $\boldsymbol{\mu}$ gradually becomes an obstacle preventing the agent from exploring better actions. The agent is constrained within the suboptimal action set provided by $\boldsymbol{\mu}$, which affects the optimality of decisions. To address this issue, we propose a dynamic action mask strategy that dynamically adjusts the value of $N_A$ during training, thereby gradually relaxing the constraints of $\boldsymbol{\mu}$ on the agent's decision-making, noted as SDAM.

\subsection{Proposed Q-SDAM Algorithm}

    \begin{figure}[ht]
        \centering
        \includegraphics[width=.8\linewidth]{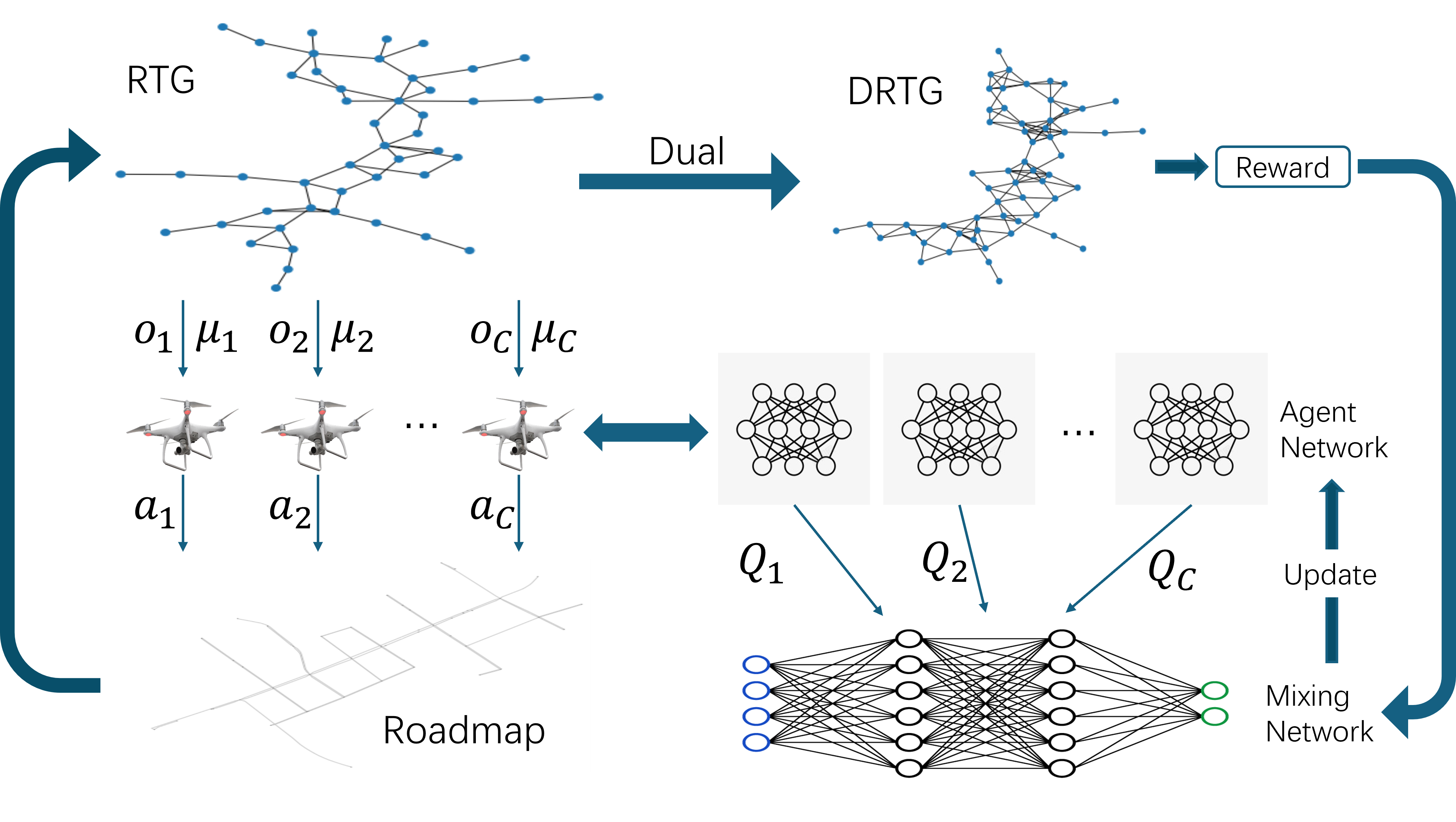}
        \caption{Architecture of Q-SDAM}
        \label{fig:method_structure}
    \end{figure}

    The architecture of Q-SDAM is illustrated in Fig. \ref{fig:method_structure}. Since the training process of the DRL algorithm incorporates operations from the execution phase, the training and execution process of Q-SDAN is presented in Algorithm \ref{algo:train}. Q-SDAM adopts a Centralized Training with Decentralized Execution (CTDE) architecture. At the start of each training round, the value of $N_A$ is linearly adjusted based on the current training progress to implement the dynamic mask strategy (Line 3). During mission execution, each UAV interacts with the environment to obtain observations, calculates the action mask based on the action scoring function defined in Eq. \eqref{equ:score_func}, and subsequently selects and executes an action based on the action mask (Lines 5-10). In the training phase, the UAVs store transitions into the replay buffer (Lines 14-17), and at regular intervals, samples are drawn from the replay buffer to update the network parameters (Lines 19-22).

    \begin{algorithm}[ht]
        \SetAlgoLined
        \caption{Training \& Execution of Q-SDAM}
        \label{algo:train}
        \KwIn{RTG $\mathcal G$, UAV set $\mathcal U$, training episode num $S$, time slot num $T$.}
        Initialize replay buffer $\mathcal B$\;
        \For{s = 1 \rm{\textbf{to}} S}{
            $N_A = (s / S) \cdot |\mathcal A|$\;
            \For{t = 1 \rm{\textbf{to}} T}{
                Obtain $s_t, o_t$ from $\mathcal B$\;
                \ForEach{$U_m$ \rm{\textbf{in}} $\mathcal U$}{
                    Obtain action mask $\mathcal \mu_{m}(t)$ based on $f(\cdot)$\;
                    $a_{m}(t) = \varepsilon\text{-greedy}(o_{m}(t),\ \mu_{m}(t))$\;
                    $U_m$ executes $a_{m}(t)$\;
                }
                \ForEach{$U_m$ \rm{\textbf{in}} $\mathcal U$}{
                    $U_m$ obtains $o_{m}(t + 1)$\;
                }
                Obtain joint action $a_t$ from $a_{m}(t)$\;
                Obtain joint observation $o_{t + 1}$ from $o_{m}(t + 1)$\;
                Obtain the global state $s_{t + 1}$ and reward $r_t$\;
                Add transition $(s_t, o_t, a_t, r_t, s_{t + 1}, o_{t + 1})$ to $\mathcal{B}$\;
            }
            \If{Sampling conditions are met}{
                Sample from $\mathcal{B}$ and calculate the loss function\;
                Update the network parameters\;
            }
        }
    \end{algorithm}
\section{Performance Evaluation}\label{sec:performance_evaluation}

    To verify the performance of the proposed Q-SDAM algorithm, we construct a simulation system based on the real-world dataset from \cite{DataSet} and conducted comparative experiments and ablation experiments. 

\subsection{Experiment Setup}

\begin{figure}[ht]
        \centering
        \includegraphics[width=.9\linewidth]{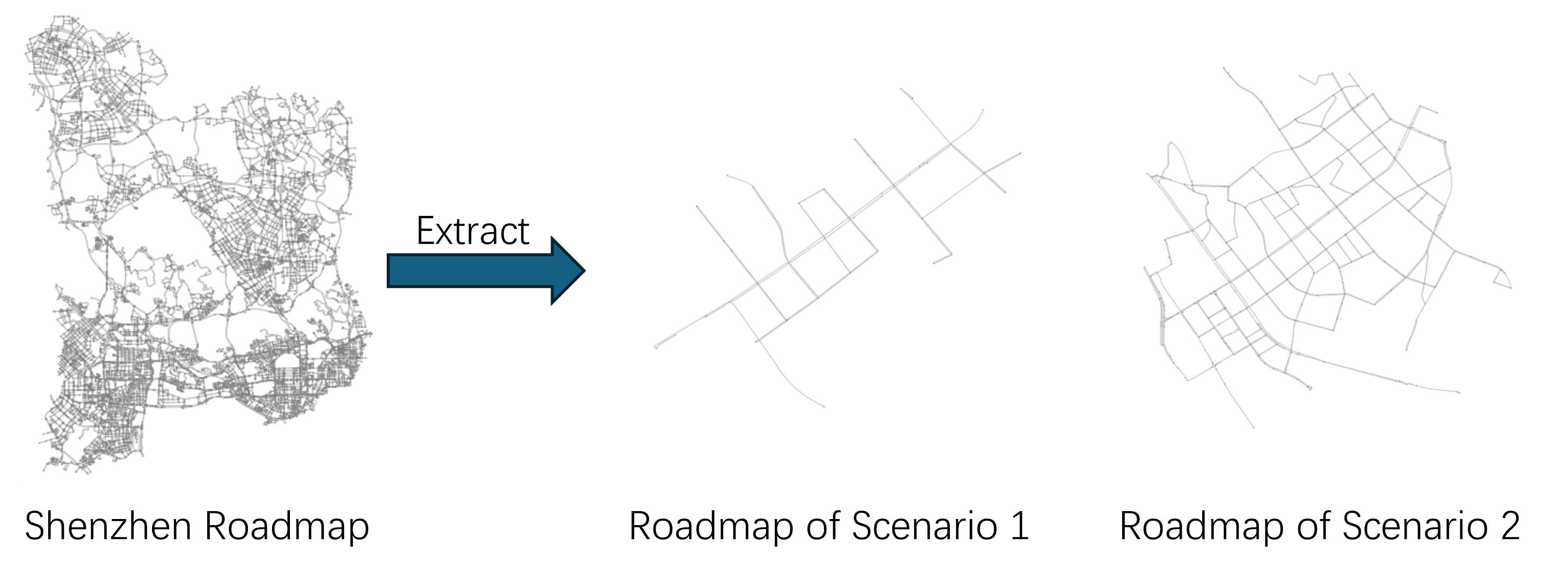}
        \caption{Roadmap of Scenarios}
        \label{fig:example_roadmap}
    \end{figure}

    The dataset from \cite{DataSet} is built using data collected from over 3,000 traffic cameras in two cities in China (Shenzhen and Jinan), forming a large-scale and structured city-level vehicle trajectory dataset, and provides urban maps in topological form covering approximately 5 million vehicle trajectory data points over a 4-day time span. We reduce the original dataset to generate two experimental roadmaps of different scales, denoted as Scenario 1 (small-scale) and Scenario 2 (large-scale), based on the data of Shenzhen on April 16, 2021. These roadmaps are used to conduct comparative experiments and verify the performance of the proposed Q-SDAM algorithm. The scales of the roadmaps are shown in Fig. \ref{fig:example_roadmap}. All experiments were conducted on a Fedora 42 Server equipped with 2 $\times$ Intel(R) Xeon(R) Gold 5318Y @ 3.40 GHz and 4 $\times$ NVIDIA GeForce RTX 3080 10GB.

    We evaluate the performance of the proposed Q-SDAM through comparisons with the following algorithms:

    \begin{itemize}
        \item Q-SAM: Based on Q-SDAM, with $N_A$ set as a fixed value.
        \item DISCOUNT: A DRL-based multi-UAV trajectory planning algorithm for VANET proposed in \cite{discount}.
        \item $\mu$-Greedy: UAVs always select the action with the highest Action Score.
        \item Random: A baseline algorithm where UAVs randomly select target positions.
    \end{itemize}

    Besides, to measure the performance of each algorithm in the task of multi-UAV-aided VANET enhancement, we compare the following performance metrics: the average number of vehicles in c-components ($M_C$), the average size of c-components ($M_E$), and the average flight distance of UAVs per time slot ($M_F$). Among these, $M_C$ and $M_E$ correspond to the optimization objective $\mathcal O_1$ proposed in \eqref{equ:O_1}, where high values are preferable. $M_F$ corresponds to the optimization objective $\mathcal O_2$ proposed in \eqref{equ:O_2}, which should be minimized.

\subsection{Numerical Results and Discussions}

\begin{figure}[ht]
        \centering
        \includegraphics[width=.7\linewidth]{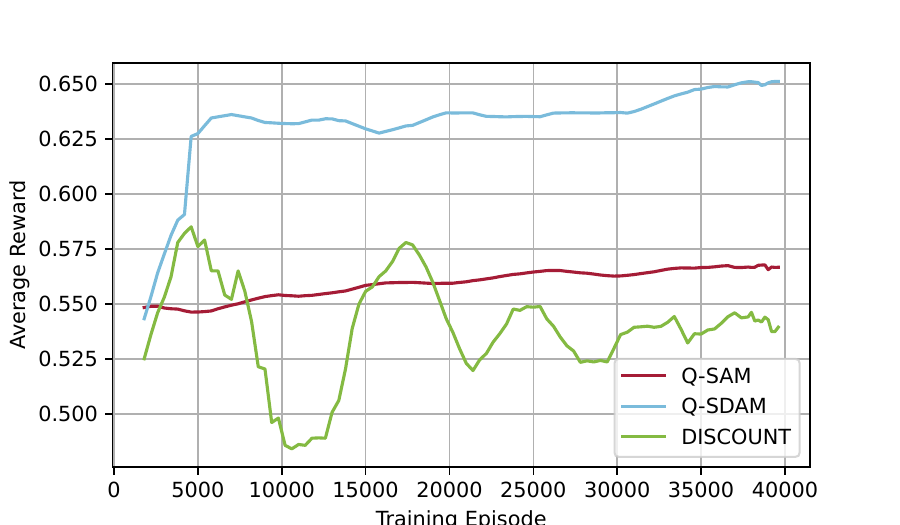}
        \caption{Converge of Intelligent Algorithms}
        \label{fig:train_reward}
    \end{figure}

    Fig. \ref{fig:train_reward} illustrates the convergence process of each intelligent algorithm. First, benefiting from the design of SDAM, the process of UAVs exploring the environment is effectively guided, which significantly accelerates the convergence speed of the proposed Q-SDAM and Q-SAM. In contrast, due to the excessively large action space without targeted design, the training process of DISCOUNT exhibits severe oscillation, and its convergence speed is much slower than that of Q-SDAM. Second, as mentioned earlier, thanks to the dynamic action mask strategy, Q-SDAM can break through local optima and obtain the optimal policy after training. However, Q-SAM is restricted by the suboptimality of $\boldsymbol{\mu}$, leading UAVs to fall into local optima and unable to optimize their action policies even after extensive episode training.

    \begin{figure}[ht]
        \centering
        \subfloat[Scenario 1]{
            \includegraphics[width=.7\linewidth]{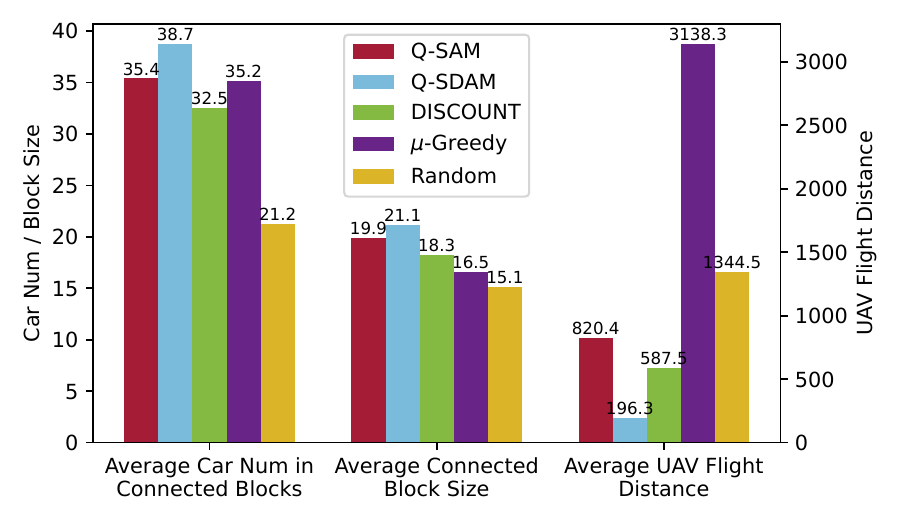}
        } \\
        \subfloat[Scenario 2]{
            \includegraphics[width=.7\linewidth]{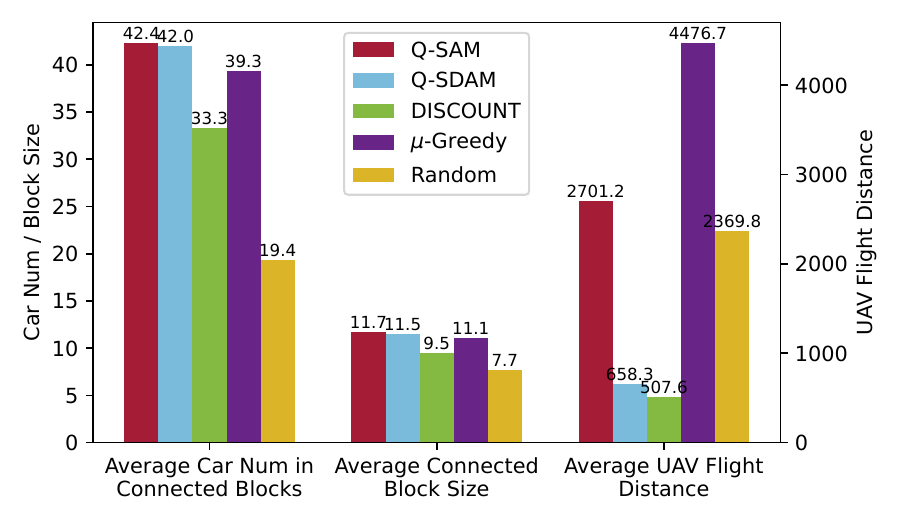}
        }
        \caption{Performance Comparison of Algorithms}
        \label{fig:algo_cmp}
    \end{figure}
    
     Fig. \ref{fig:algo_cmp} shows the performance of each algorithm in Scenario 1 and Scenario 2. In Scenario 1, Q-SDAM outperforms the comparison algorithms significantly in all metrics: compared with DISCOUNT, it increases $M_C$ by 18.2\% and $M_E$ by 15.3\%, while its $M_F$ is reduced by 66.6\%. In Scenario 2, the $M_C$ and $M_E$ of Q-SDAM are slightly inferior to those of Q-SAM (by approximately 1\%), but its $M_F$ is reduced by 75.6\%. The $M_F$ of Q-SDAM is 8\% worse than that of DISCOUNT, but its $M_C$ is 26.1\% higher than that of DISCOUNT. This indicates that the proposed Q-SDAM, under the guidance of SDAM, conducts sufficient exploration of the environment and learns the optimal policy. Notably, benefiting from the design of the dynamic action mask mechanism, Q-SDAM avoids falling into local optima like Q-SAM. The action policy of Q-SAM is close to that of $\mu$-Greedy, showing good performance in $M_C$ and $M_E$ but poor performance in $M_F$.

    \begin{figure*}[htb]
        \centering
        \subfloat[$M_C$]{
            \includegraphics[width=.3\linewidth]{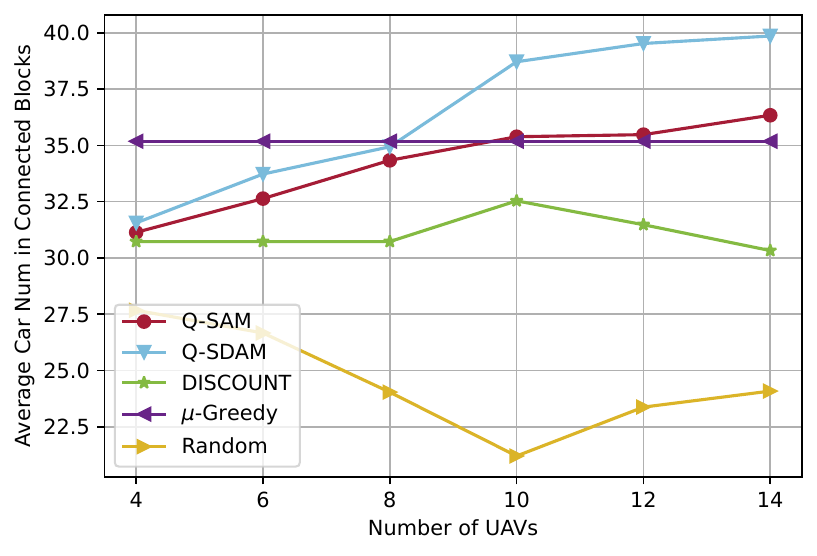}
        }
        \subfloat[$M_E$]{
            \includegraphics[width=.3\linewidth]{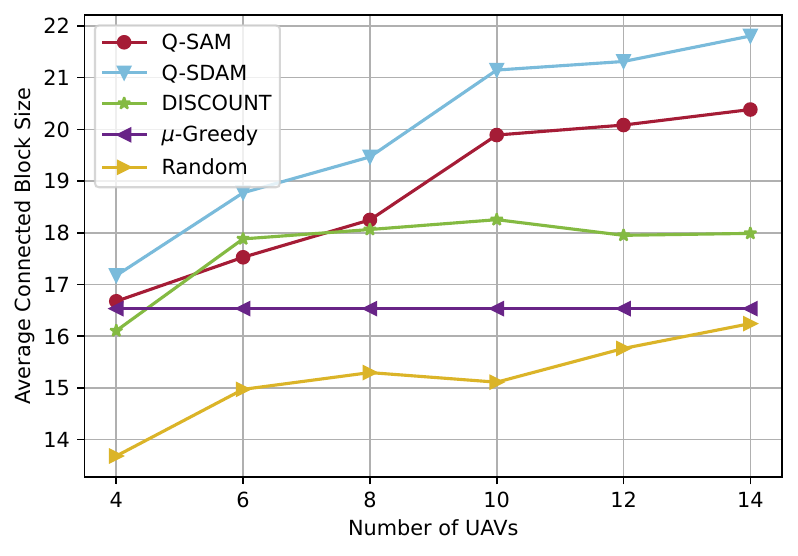}
        }
        \subfloat[$M_F$]{
            \includegraphics[width=.3\linewidth]{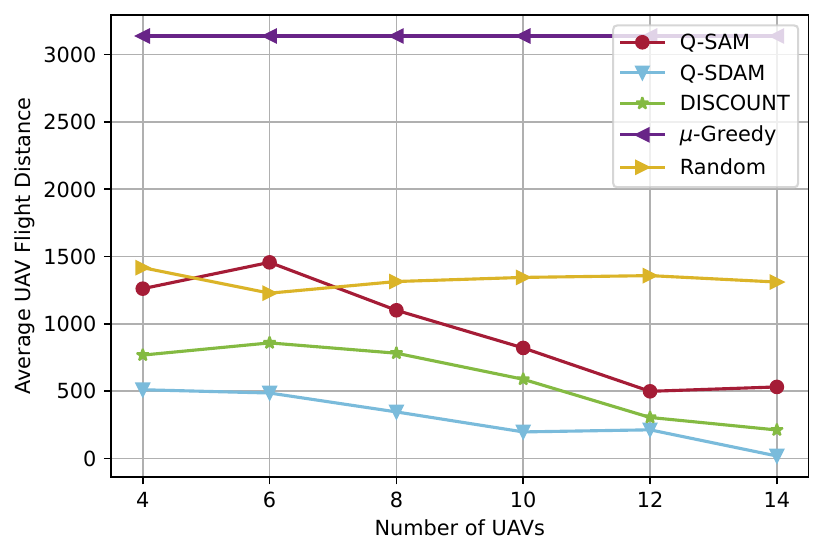}
        }
        \caption{Performance Comparison with Varying Number of UAVs}
        \label{fig:arg_cmp}
    \end{figure*}

    To verify the robustness of the proposed algorithm under different scenarios, experiments were conducted with varying numbers of UAVs, as shown in Fig. \ref{fig:arg_cmp}. As the number of UAVs increases, the performance metrics of all algorithms generally show an upward trend. The proposed Q-SDAM can effectively handle the sudden expansion of the decision space caused by the increase in the number of UAVs, demonstrating excellent robustness. In contrast, due to the lack of targeted design, the performance of DISCOUNT begins to decline instead of improving when the number of UAVs exceeds 10. Notably, since the design of SDAM only relies on the UAV's position in the previous time slot and the vehicle distribution, the performance metrics of $\mu$-Greedy remain unchanged with variations in the number of UAVs. Regarding Random (used as the baseline), the increase in the number of UAVs leads to more empty c-components, resulting in an increase in its $M_E$ metric but a decrease in its $M_C$ metric.

\section{Conclusion}\label{sec:conclusion}
    
    This paper has studied multi-UAV deployment to enhance VANET connectivity in urban environments. To tackle multi-objective optimization and large action spaces, we proposed the Q-SDAM algorithm, which employed a score-based dynamic action mask to guide MADRL training for intelligent UAV deployment. Experiments on real-world datasets have shown that Q-SDAM converged quickly , effectively avoided local optima, and outperformed existing algorithms in connectivity and energy efficiency. Future work will refine action mask generation and incorporate vehicle dynamics to further improve scalability in urban scenarios.

\bibliographystyle{IEEEtran}
\bibliography{IEEEabrv, reference}
\end{document}